\documentclass[aps,prl,twocolumn,superscriptaddress]{revtex4}
\usepackage{graphicx}
\usepackage{amssymb}
\usepackage{amsmath}
\usepackage{graphicx}
\usepackage{epsfig}
\usepackage{color}

\begin{document}

\title{Universal behavior of the thermal Hall conductivity}
\author{Yi-feng Yang}
\email[]{yifeng@iphy.ac.cn}
\affiliation{Beijing National Lab for Condensed Matter Physics and Institute of Physics,
Chinese Academy of Sciences, Beijing 100190, China}
\affiliation{School of Physical Sciences, University of Chinese Academy of Sciences,
Beijing 100190, China}
\affiliation{Songshan Lake Materials Laboratory, Dongguan, Guangdong 523808, China}
\author{Guang-Ming Zhang}
\email[]{gmzhang@tsinghua.edu.cn}
\affiliation{State Key Laboratory of Low-Dimensional Quantum Physics and Department of
Physics, Tsinghua University, Beijing 100084, China}
\affiliation{Frontier Science Center for Quantum Information, Beijing 100084, China}
\author{Fu-Chun Zhang}
\email[]{fuchun@ucas.ac.cn}
\affiliation{Kavli Institute for Theoretical Sciences and CAS Center for Topological
Quantum Computation, University of Chinese Academy of Sciences, Beijing
100190, China}
\date{\today }

\begin{abstract}
We report theoretical and experimental analyses of the thermal Hall conductivity in correlated systems. For both fermionic and bosonic excitations with nontrivial topology, we show that at ``intermediate" temperatures, the thermal Hall conductivity exhibits an unexpected universal scaling with a simple exponential form. At low temperatures, it behaves differently and reflects the spectral properties of underlying excitations. Our predictions are examined as examples in two prototype compounds, the quantum paraelectric SrTiO$_3$ and the spin-liquid compound RuCl$_3$. The experimental data can be largely covered by our proposed minimal phenomenological model independent of microscopic details, revealing dominant bosonic contributions in SrTiO$_3$ and gapped fermionic excitations in RuCl$_3$. Our work establishes a phenomenological link between microscopic models and experimental data and provides a unified basis for analyzing the thermal Hall conductivity in correlated systems over a wide temperature region.
\end{abstract}

\maketitle

\textbf{Introduction} - The thermal Hall conductivity probes topologically nontrivial excitations in a correlated insulator. In particular, the recent observation of ``giant" thermal Hall conductivity in cuprates \cite{Grissonnanche2019} has stimulated intensive
interest in understanding its underlying mechanism and potential topological excitations \cite{Han2019,Scheurer2018,Sachdev2019,DHLee2019,Tao2019}, although the issue is still highly controversial. Previously, the thermal Hall conductivity has also been observed in spin liquid systems such as RuCl$_3$ but the signal is ten times smaller \cite{Kasahara2018,Kasahara2018b,Hentrich2019}. By contrast, a twice larger effect was lately reported in SrTiO$_3$ and attributed to phonons \cite{Li2019}. While the cuprate experiment has yet to be confirmed by other groups, it has posed an urgent challenge for our theoretical understanding of this unusual phenomena in general. Complicated microscopic models have since been developed but involve various adjusting parameters that prevent straightforward comparisons with experiments \cite{Han2019,Scheurer2018,Sachdev2019,DHLee2019,Tao2019,Gao2019,Gao2019b,Tang2019,Chen2019,Yang2019,Yang2019b,Zhang2020}. So far, all scenarios remain preliminary and no well-established connection has been known between existing experimental observations and theoretical proposals. From the viewpoint of a more complete theory of the thermal Hall conductivity, there is obviously a missing link between microscopic models and experimental data.

\begin{figure}[b]
\centerline{{\includegraphics[width=.45\textwidth]{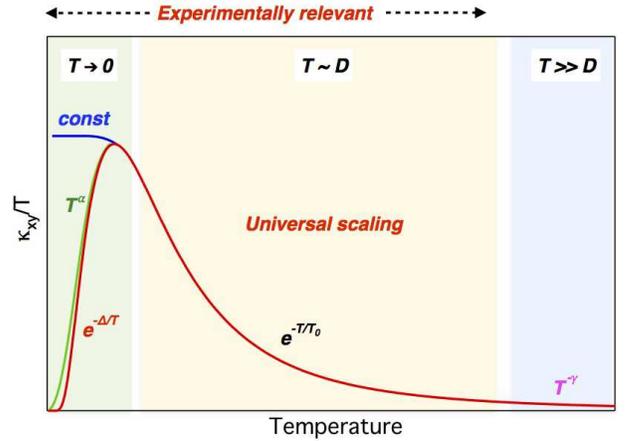}}}
\caption{Sketch of three temperature regions for the thermal Hall conductivity: the low-temperature region with saturating ($const$), power-law ($T^\alpha$), or activation ($e^{-\Delta/T}$) behaviors that reflect different spectral properties of underlying excitations; the intermediate temperature region with a universal scaling of the exponential form ($e^{-T/T_0}$) for $T\sim D$, where $D$ represents the characteristic energy for topologically nontrivial excitations; the high $T$ limit with $T^{-\gamma}$ scaling, where $\gamma=1$ for bosons and $\gamma=3$ for fermions. Only the former two regions are relevant in experiment.}
\label{fig1}
\end{figure}

To fill in this gap, we notice that, in resemblance of the Fermi liquid theory, instead of starting with a material-dependent Hamiltonian for individual systems, it is possible to make minimal assumptions on underlying fermionic or bosonic excitations and look for generic features (similarities and differences) to be compared with existing experiments. This allows us to identify if and to what extent the data may be understandable without resorting to microscopic details. As a result, we find that the experiment can be well covered by two temperature regions: a universal exponential scaling of the thermal Hall conductivity at intermediate temperatures and different low-temperature signatures (saturation, power-law or activation) for different types of charge neutral excitations. These are summarized in Fig.~\ref{fig1}, showing the above two regions and a third region in the high $T$ limit. The latter is, however, irrelevant in experiment. Our results are then confirmed as examples in two prototype compounds, the quantum paraelectric SrTiO$_3$ and the spin-liquid compound RuCl$_3$, revealing their different elementary excitations and similar universal temperature scaling.

\textbf{Theoretical analyses} - Microscopic Hamiltonians typically include complicated many-body interactions, but their low-energy physics may often be described by well-defined fermionic or bosonic quasiparticles such as spinons, magnons, or phonons. For charge neutral excitations with nontrivial band topology, a thermal Hall conductivity can arise following the basic Mott-like formula \cite{Qin2011,Qin2012,Wang2009,Matsumoto2011,Matsumoto2014},
\begin{equation}
\frac{\kappa_{xy}}{T}=\frac{1}{T^{2}}\int (\epsilon-\mu)^{2}\left(-\frac{\partial n}{\partial \epsilon }\right) \sigma _{xy}(\epsilon)d\epsilon,
\end{equation}
where $n(\epsilon) =[e^{\beta(\epsilon -\mu)}\pm1]^{-1}$ denote the Fermi-Dirac (+) or Bose-Einstein (-) distribution functions and $\mu$ is the chemical potential. For intrinsic contribution, $\sigma _{xy}(\epsilon )$ is the zero-temperature Hall conductivity, which is associated with the Berry curvature through the well known Thouless-Kohmoto-Nightingale-Nijs (TKNN) formula \cite{TKNN}, 
\begin{equation}
\sigma_{xy}(\epsilon)=\sum_{nks}\mathcal{B}_{nks}\theta (\epsilon-E_{nks}),
\end{equation}
where the subscripts $n$, $k$, $s$ denote the band, crystal momentum, and (pseudo)spin indices, respectively. $\mathcal{B}_{nks}$  and $E_{nks}$ represent the corresponding Berry curvature and quasiparticle dispersion. The exact forms of these quantities are often complicated and depend on the details of the microscopic Hamiltonian. For extrinsic contribution, $\sigma _{xy}(\epsilon )$ has a more complicated form but is still related to the Berry curvature. In both cases, we may avoid the complication and focus on generic features of the thermal Hall conductivity by considering the linear-in-field approximation, 
\begin{equation}
\sigma_{xy}(\epsilon )\approx H\left.\frac{\partial\sigma_{xy}}{\partial H}\right|_{H=0}\equiv H\mathcal{B}(\epsilon) .
\end{equation}
To illustrate the idea, we first consider a simplest example. For spinons with only the Zeeman coupling, we may have $\mathcal{B}(\epsilon)=\sum_{nk}\mathcal{B}_{nk}\delta(\epsilon -E_{nk})$ if $\mathcal{B}_{nks}=s\mathcal{B}_{nk}$ and $E_{nks}=E_{nk}-sH$ with $s=\pm$ \cite{Han2019}. Thus $\mathcal{B}(\epsilon)$ represents tentatively the effective density of the Berry curvature as a function of energy. Evidently, $\kappa_{xy}(T)/TH$ is fully determined by $\mathcal{B}(\epsilon)$, which is a convolution of the Berry curvature and the carrier bands or density of states (DOS). The latter reminds us that the exact form of the DOS is often not essential and the key physical properties may be obtained with simple assumptions, e.g., constant DOS for gapless fermions \cite{HewsonBook}. Following this line of thought, we make the attempt to extend similar assumptions to $\mathcal{B}(\epsilon)$ and examine if a generic understanding of the thermal Hall conductivity can be achieved without resorting to the band details. More generally, the magnetic field may also couple to the orbital moment, resulting in a more involved form of $\mathcal{B}(\epsilon)$ \cite{Xiao2005,Thonhauser2005,Shi2007}. For phonons, $\mathcal{B}(\epsilon)$ comes  from the linear-in-field term of the Berry curvature \cite{Qin2012}. But regardless of all these details, nontrivial topological effect typically only exists within a finite energy window ($D$), out of which $\mathcal{B}(\epsilon)$ can be safely set to zero. We will see that this simple fact, combined with the small magnitude of $D$, yields a key experimental feature of the thermal Hall conductivity in all correlated insulators.

\begin{figure}[t]
\centerline{{\includegraphics[width=.5\textwidth]{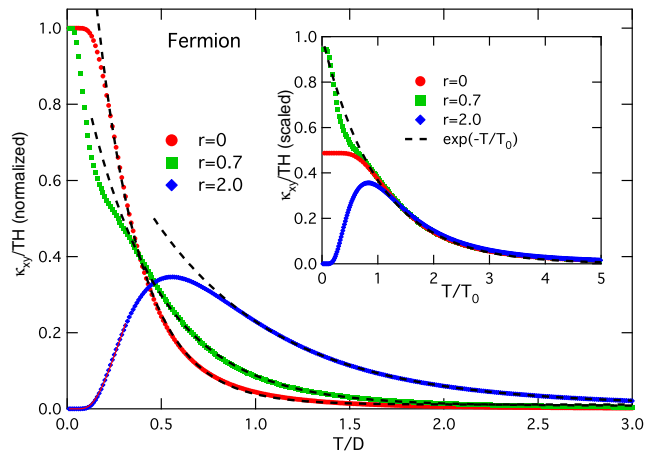}}}
\caption{The calculated $\protect\kappa_{xy}/TH$ as a function of $T/D$ for the fermionic model with different values of $r=\mu/D$, showing saturation for $|r|<1$ and a maximum for $|r|>1$ at low temperatures. We find excellent exponential temperature dependence at relatively higher temperatures. The dots are numerical results from the Mott-like formula and the dashed lines are fitting curve of a simple exponential form. The inset shows the collapse of the calculated thermal Hall conductivity with respect to $T/T_{0}$ after proper rescaling for different $r$.}
\label{fig2}
\end{figure}

We first study the case of fermionic excitations. To simplify the discussion, we assume a flat ``Berry density",
\begin{equation}
\mathcal{B}(\epsilon )=\left\{
\begin{array}{cc}
\mathcal{B}_{0} & -D\leq \epsilon \leq D \\
0 & \text{otherwise}
\end{array}
\right.,
\end{equation}
where $\mathcal{B}_{0}$ is a constant and $[-D, D]$ denotes the energy window with nontrivial band topology. A close example can be found in Ref.~\protect\cite{Han2019}, which includes two spinon bands with opposite Berry curvatures. The thermal Hall conductivity is then given by
\begin{equation}
\frac{\kappa_{xy}}{TH}=\mathcal{B}_{0}\int_{-(1+r)/t}^{(1-r)/t}{d}x\frac{x^{2}}{\cosh^{2}\left(x/2\right)},
\label{Eq4}
\end{equation}
where the ratio $r=\mu/D$ is the relative chemical potential and $t=T/D$ is the dimensionless temperature. At zero temperature, the integral can be evaluated exactly, yielding $\kappa_{xy}/TH=2\pi ^{2}\mathcal{B}_{0}/3$ for $|r|<1$ (gapless) and zero for $|r|>1$ (gapped). The thermal Hall conductivity with lowering temperature is therefore expected to saturate ($|r|<1$) or diminish as $e^{-\Delta/T}$ ($|r|>1$), where $\Delta \propto (|r|-1)D$ is the effective activation gap of the fermions. In the high $T$ limit ($T\gg D$), one has $\kappa_{xy}/TH\propto T^{-3}$.

In between, for intermediate temperature of the order of $D$, which is in fact most relevant in experiment, we find a different scaling,
\begin{equation}
\frac{\kappa_{xy}}{TH}\sim  e^{-T/T_{0}},
\end{equation}
where $T_{0}$ represents a temperature scale characterizing the decay of the thermal Hall conductivity with increasing temperature. The numerical results are shown in Fig.~\ref{fig2} in comparison with the exponential fit. A closer inspection indicates that the deviation from the exponential fit appears also roughly below $T_{0}$, which not only marks the characteristic temperature of the universal scaling but also separates the different regions of the thermal Hall effect. To see this more clearly, we rescale the curves with the fitting parameters. As shown in the inset of Fig.~\ref{fig2}, they all collapse onto a single exponential curve above $T_{0}$. The properties of $\kappa_{xy}$ over a wide temperature range are then controlled entirely by two parameters: the temperature scale $T_{0}\propto D$ and the average Berry density $\mathcal{B}_0$ that determines the amplitude and sign of the thermal Hall signal.

The thermal Hall conductivity contributed by bosons, in particular phonons, might involve other complicated effects. In ferroelectric insulators, extrinsic skew scattering was proposed to govern the thermal Hall effect \cite{Chen2019}. In this case, the Mott-like formula can still be applied to describe the temperature gradient of phonon excitations ($\partial n/\partial\epsilon$), but $\mathcal{B}(\epsilon)$ is no longer given by the TKNN formula and becomes an effective quantity induced by the phonon skew scattering. For gapless bosons (acoustic phonons or magnons), the chemical potential is zero. The bosonic $\mathcal{B}(\epsilon)$ may be assumed to take the power-law form at low energies,
\begin{equation}
\mathcal{B}(\epsilon )=\left\{
\begin{array}{cc}
\mathcal{B}_{0}\epsilon^\alpha & 0\leq \epsilon \leq D \\
0 & \text{otherwise}
\end{array}
\right.,
\end{equation}
where $\alpha$ is a constant exponent and $D$ is the cutoff energy that constraints the nontrivial topology of the bosonic states. Detailed examples may be found in the literature for acoustic phonons in magnetic solids ($\alpha=2$) \cite{Qin2012} or nonmagnetic insulators ($\alpha=1$) \cite{Saito2019}. We have
\begin{equation}
\frac{\kappa_{xy}}{TH}=\mathcal{B}_{0}T^\alpha\int_{0}^{1/t}dx\frac{x^{2+\alpha}}{\sinh^{2}\left( x/2\right)},
\end{equation}
whose temperature evolution only depends on the values of the exponent $\alpha$. 

\begin{figure}[t]
\centerline{{\includegraphics[width=.5\textwidth]{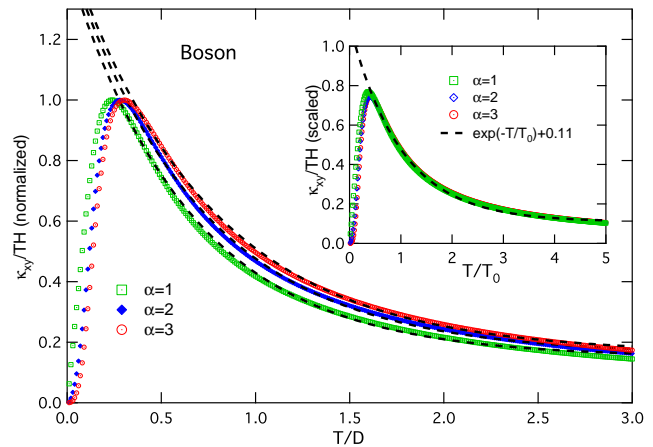}}}
\caption{The calculated $\protect\kappa_{xy}/TH$ as a function of $T/D$ for the bosonic model with different values of $\alpha$, showing a maximum for $\alpha>0$ at low temperatures and good exponential temperature dependence at relatively higher temperatures. The dots are numerical results and the dashed lines are the exponential fit. As shown in the inset, all curves of different $\alpha$ collapse onto a single exponential function of $T/T_{0}$.}
\label{fig3}
\end{figure}

At low temperatures, this yields $\kappa_{xy}/TH\propto T^\alpha$, namely, a power-law temperature dependence. The exponent $\alpha$ can thus be measured experimentally and reflects some information on the microscopic origin of the bosonic excitations. Once again, at intermediate temperatures ($T\sim D$), our numerical results in Fig.~\ref{fig3} show similar exponential temperature dependence, $e^{-T/T_0}$, for all values of $\alpha$. With proper rescaling, as plotted in the inset of Fig.~\ref{fig3}, all data collapse onto a single exponential curve, albeit with a constant offset that originates from the difference between Bose and Fermi distributions, namely, $\cosh(x)\rightarrow 1$ and $\sinh(x)\rightarrow x$ for $x\rightarrow 0$ in their respective Mott-like formulas. Such a distinction has also a consequence in the high $T$ limit. For bosons, we have instead $\kappa_{xy}/TH\propto T^{-1}$ for $T\gg D$. The separation of low and intermediate temperature regions is not universal in reality and also no longer associated with $T_{0}$. For gapped bosons such as optical phonons (not shown), one expects the thermal activation behavior, $e^{-\Delta/T}$, at low $T$, but similar exponential scaling at intermediate $T$.

We emphasize again that the exponential scaling is not a high $T$ limit but an approximation for intermediate $T$ of the order of $D$. To see how it appears, we take as an example $r=0$ and rewrite the right hand side of Eq.~(\ref{Eq4}) as $2\mathcal{B}_0\int^\infty_t dy e^{-2g(y)}$, whose temperature derivative gives $-2\mathcal{B}_0 e^{-2g(t)}$, with $g(t)=2\ln t+\ln\cosh(1/2t)$. This defines an inflection point $t^*$ from $g^{\prime\prime}(t^*)=0$, such that for $t$ around $t^*$, or more precisely $(t-t^*)^2\ll 6|g^\prime(t^*)/g^{\prime\prime\prime}(t^*)|$, we have the expansion, $e^{-2g(t)}\approx e^{-2g(t^*)-2(t-t^*)g^\prime(t^*)}\propto e^{-t/t_0}$, where $t_0=1/2g^\prime(t^*)$ and the higher order terms are neglected. The integral then yields the exponential scaling around $t^*$ up to a constant offset. Note that $g(t)$ has a maximal slope at $t^*$. Hence the exponential scaling appears where $\kappa_{xy}/TH$ decays most rapidly and has a minimal $t_0$. For $r=0$, the above analysis gives $t^*\approx 0.49$ and $t_0\approx 0.2$ for the exponential temperature dependence as shown in Fig.~\ref{fig2}. The exponential scaling thus appears over a finite temperature window of the order of $D$ around the inflection point $t^*D$ (or twice $T_0$) but breaks down in the limit $T/D\gg1$ where it is replaced by a different scaling, $\kappa_{xy}/TH\propto T^{-\gamma}$, with $\gamma=3$ for fermions and $\gamma=1$ for bosons. The derivation can be extended to other cases. Introducing a general but regular $\mathcal{B}(\epsilon)$ within the energy window $D$ might complicate the derivation but will not invalidate the linear-in-$T$ approximation on the exponent. The existence of the inflection point $t^*$ is a property of the intermediate temperature region connecting the peculiar low and high $T$ limits of $\kappa_{xy}/TH$. 

Putting together, our theoretical analyses reveal two generic features of the thermal Hall conductivity of the minimal model for both fermionic and bosonic excitations. First, its low temperature saturating, power-law, or activation behavior reflects the spectral properties of underlying excitations. Second, it has a universal exponential scaling at relatively higher (or intermediate) temperatures ($T\sim D$). This exponential form appears over a finite temperature range above $T_0$ where $\kappa_{xy}$ is given by the thermal average of all topological excitations within a narrow energy window ($D\propto T_0$). The highest $T$ limit ($T\gg D$) is, however, often beyond the relevant temperature range where the discrepancy from the exponential scaling becomes irrelevant or indiscernible in experiment. The three different regions are sketched in Fig.~\ref{fig1} and independent of the model details.

\textbf{Experimental analyses} - Next we examine existing experiments in real materials RuCl$_3$ \cite{Kasahara2018,Kasahara2018b,Hentrich2019} and SrTiO$_3$ \cite{Li2019} to see if they may be understood by the minimal model or to what extend further band details will be needed in order to explain the data.

\begin{figure}[t]
\centerline{{\includegraphics[width=.48\textwidth]{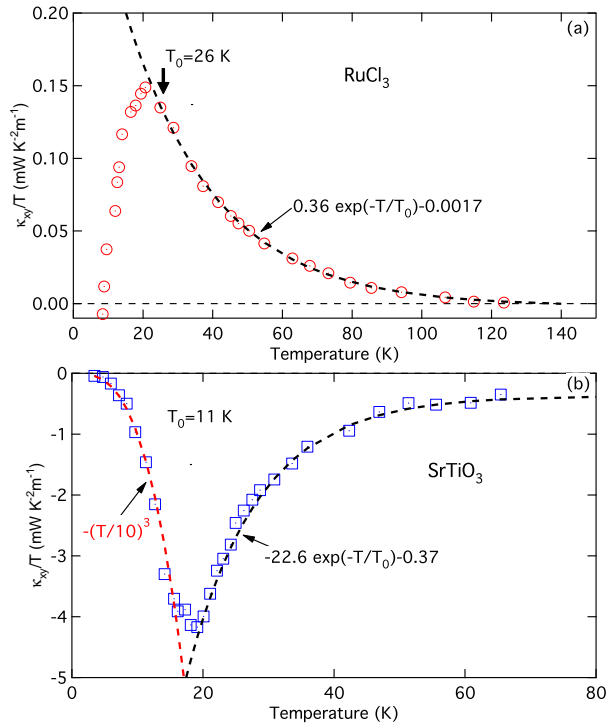}}}
\caption{Original data of the thermal Hall conductivity reproduced for (a) $\alpha$-RuCl$_3$ \cite{Hentrich2019,Grissonnanche2019} and (b) SrTiO$_3$ \cite{Li2019}. The dots represent the experimental data and the dashed lines denote the theoretical fit with an exponential or power-law function.}
\label{fig4}
\end{figure}

The experimental data for SrTiO$_3$ and RuCl$_3$ are replotted in Fig.~\ref{fig4}. In both cases, we see good exponential fits at higher  temperatures, in excellent agreement with the prediction of the minimal model at intermediate temperatures. A couple of fitting details may reveal the difference between the two compounds. First, the exponential fit yields a negligible offset for RuCl$_3$ but a relatively large one for SrTiO$_3$. Second, the deviation from the fit appears around $T_0\approx 26$ K for RuCl$_3$ but far away from $T_0\approx11\,$K for SrTiO$_3$. At lower temperatures, $\kappa_{xy}/T$ follows the power law with $\alpha=3$ for SrTiO$_3$, suggesting a bosonic (phonon) $\mathcal{B}(\epsilon)\propto \epsilon^3$ for small $\epsilon$. This exponent is, however, different from the general expectation for the intrinsic Hall conductivity of acoustic phonons \cite{Qin2012}, but is consistent with the proposal of an extrinsic skew scattering mechanism \cite{Chen2019}. For RuCl$_3$, the low-$T$ suppression of $\kappa_{xy}/T$ indicates gapped fermionic (spinon) excitations, as suggested by the long-range magnetic order below 8 K. Unfortunately, we cannot fit the data with the activation function due to the intervene of the magnetic order. Nevertheless, it is clear that following our theory, these differences imply that the responsive excitations are fermionic (spinons) in RuCl$_3$ and bosonic (phonons) in SrTiO$_3$, consistent with usual expectations.

The experimental data in both types of compounds may therefore be well understood by our phenomenological theory in two temperature regions. The excellent agreement for both compounds confirm the validity of our simple model. Since $T_0\propto D$, the exponential scaling observed below around 100 K in both compounds implies that nontrivial topological excitations only have an effect within a narrow energy window ($D$) of few meV. Other microscopic details seem irrelevant in explaining the observed temperature scaling. We should note that a realistic system may often have several bands with different topological properties. Our model then approximates each band (or energy window) individually. The total  thermal Hall conductivity is the sum of all, but its low temperature property will be dominated solely by the one near the Fermi energy. It is easy to see that the presence of multiple bands will not change the generic features. On the other hand, going beyond the free quasiparticle picture, residual many-body interactions might cause a finite lifetime of the elementary excitations and further suppress the thermal Hall conductivity, possibly causing an even faster decay with increasing temperature and hence a reduced $T_0$.

The overall magnitude of the thermal Hall conductivity can also be understood roughly from the perspective of the minimal model. The zero temperature limit of $\kappa_{xy}/T$ reflects the Berry curvature of underlying excitations at the Fermi energy, but its magnitude from the high temperature fit (which is the case for most experiments) is an effect of thermal average. For phonons as in SrTiO$_3$, the latter might involve complicated scattering mechanism and is therefore not intrinsic. For gapless fermions, the magnitude of $\kappa_{xy}/T$ continues to increase with lowering temperature and should eventually saturate to the order of the total Chern number of filled fermionic bands in unit of $k_B^2/h$. For gapped fermions, the low temperature limit is zero and there may exist both valence and conduction bands with opposite Berry curvatures below and above the Fermi energy \cite{Han2019}. As a result, the high temperature $\kappa_{xy}/T$ could often suffer from potential cancellation of the opposite contributions of the two bands. This might explain the relatively small magnitude of the thermal Hall conductivity observed in RuCl$_3$. 

\textbf{Conclusions} - We have explored generic properties of the thermal Hall conductivity based on a phenomenological approach with minimal assumptions. Our detailed comparisons between theory and experiment reveal two (experimentally relevant) distinct regions of its temperature dependence: a universal exponential scaling at intermediate temperature and a saturating or power-law or activation behavior at lower temperature that reflects the spectral properties of underlying fermionic or bosonic excitations. These are confirmed in the quantum paraelectric SrTiO$_3$ and the spin-liquid compound RuCl$_3$, revealing their different elementary excitations and similar exponential temperature scaling. We have also analyzed the cuprate data \cite{Grissonnanche2019} and found excellent agreement with our proposed scaling, but to which category the carriers might belong requires more elaborate measurements at sufficiently lower temperatures. Our work establishes a phenomenological link between candidate microscopic theories and experimental data in real materials. It may be applied directly to other materials to provide useful information on potential topological excitations and the mechanism of their emergence and breakdown.
 
 \textit{Acknowledgment }- We thank Junren Shi and Tao Li for stimulating discussions. This work was supported by the National Key Research and Development Program of MOST of China (2017YFA0303103, 2017YFA0302902, 2016YFA0300300), the National Natural Science Foundation of China (11774401, 11974397, 11674278), the State Key Development Program for Basic Research of China (2014CB921203), the Strategic Priority Research Program of CAS (Grant No. XDB28000000), and the Youth Innovation Promotion Association of CAS.

\end{document}